\documentclass[12pt,preprint]{aastex}

\slugcomment{}

\shorttitle{FeLoBALs, ULIRGs, and QSOs}
\shortauthors{Farrah et al}

\begin{document}

\title{Evidence that FeLoBALs may signify the transition between an ultraluminous infrared galaxy and a quasar}


\author{D. Farrah\altaffilmark{1}}
\author{M. Lacy\altaffilmark{2}}
\author{R. Priddey\altaffilmark{3}}
\author{C. Borys\altaffilmark{4}}
\author{J. Afonso\altaffilmark{5}}

\altaffiltext{1}{Department of Astronomy, Cornell University, Ithaca, NY 14853, USA}
\altaffiltext{2}{Spitzer Science Center, California Institute of Technology, Pasadena, CA 91125, USA}
\altaffiltext{3}{Centre for Astrophysics Research, University of Hertfordshire, College Lane, Hatfield AL10 9AB, UK}
\altaffiltext{4}{Department of Astronomy and Astrophysics, University of Toronto, Toronto, Canada}
\altaffiltext{5}{Centro de Astronomia e Astrofísica, Universidade de Lisboa, Lisbon, Portugal}

\begin{abstract}
We present mid/far-infrared photometry of nine FeLoBAL QSOs, taken using the {\it Spitzer} space telescope. All nine 
objects are extremely bright in the infrared, with rest-frame 1-1000$\mu$m luminosities comparable to those of 
Ultraluminous Infrared Galaxies. Furthermore, a significant fraction of the infrared emission from many, and possibly 
all of the sample is likely to arise from star formation, with star formation rates of order several hundred solar 
masses per year. We combine these results with previous work to propose that FeLoBALs mark galaxies and QSOs in which 
an extremely luminous starburst is approaching its end, and in which a rapidly accreting supermassive black hole is in 
the last stages of casting off its dust cocoon. FeLoBAL signatures in high redshift QSOs and galaxies may thus be an 
efficient way of selecting sources at a critical point in their evolution. 
\end{abstract}


\section{Introduction}
First seen in 1967 \citep{lyn67}, Broad Absorption Line (BAL) QSOs are those objects that show broad, deep troughs in their UV and optical 
spectra, arising from resonance line absorption in gas outflowing with velocities of $\gtrsim0.1c$ \citep{wey91,ara01,hall02,rei03}.
BAL QSOs come in three subtypes depending on which absorption features are seen. High Ionization BAL QSOs (HiBALs) show absorption in Ly$\alpha$,
NV $\lambda$1240, SiIV $\lambda$1394 and CIV $\lambda$1549, and comprise about 85\%\ of the BAL population. Low Ionization 
BAL QSOs (LoBALs) contain all the absorption features seen in HiBALs, and also show absorption in MgII $\lambda$2799 
and other low ionization species, and comprise $\sim15$\% of the BAL population. Finally, a rare class of BAL 
QSO, in addition to showing all the absorption lines seen in LoBALs, also show absorption features arising from 
metastable excited levels of iron. These are termed FeLoBAL QSOs \citep{haz87,bec97,bra02,lac02}.

Efforts to explain the origin of BALs in QSOs have been ongoing since their discovery. Broadly, there are two possibilities. 
The first is that BAL QSOs are normal QSOs viewed along a particular line of sight; in this case the absorption features 
arise when an accretion disk wind encounters a high column density, ionized gas outside the broad line region \citep{mc98,pro00}. 
The gas is driven outwards via resonance line absorption, but the high column density of the gas shields it from higher energy 
photons that would otherwise completely ionize it. In this case, BAL QSOs are those QSOs viewed along a line of sight that 
coincides with the outflowing gas \citep{elv00}. The second possibility is that BAL QSOs are youthful objects, still surrounded 
by gas and dust in which the absorption takes place; in this case the BALs do not arise due to a particular line of sight 
\citep{voi93,bec97,wil99}. 

There has been significant debate over the years on the best method to find young QSOs (e.g. \citealt{san88,kaw06}), so the 
idea that some BAL QSOs may be such objects is particularly intriguing. Most attention has focused on the LoBALs as candidate 
young QSOs \citep{lip94,can01}, as the differences between the line properties of LoBAL QSOs and those of ordinary QSOs are 
hard to explain solely in terms of different relative orientations \citep{spr92}.  Nevertheless, the picture of LoBALs being 
young QSOs is controversial; for example sub-mm observations \citep{lew03,will03,pri07} show no differences between BAL QSOs 
and ordinary QSOs. Furthermore, \citet{voi93} propose a scenario in which LoBALs form via ablation of dust by UV photons in 
outflows arranged either as a thick disk or as an isotropic distribution of clouds; in this case LoBALs do fit within AGN 
orientation schemes, and this scenario is consistent with polarimetric observations \citep{sch99,ogl99,hun00}.

Recently however, evidence has mounted that FeLoBAL QSOs may be the strongest candidates for being youthful QSOs. Based on 
UV and optical spectra, \citet{hall02} suggest that FeLoBALs are young objects still surrounded by a dust cocoon. Similar 
conclusions are reached by \citet{gre02}, who also postulate that FeLoBALs may be associated with galaxy interactions. 
Further evidence comes from observations of the only two known FeLoBALs at $z<0.15$, both of which are associated with 
Ultraluminous Infrared Galaxies (ULIRGs, e.g. \citealt{far05}). Finally, the presence of winds with large spatial extents 
($\gtrsim100$pc) in some FeLoBAL QSOs \citep{dek02} provides a plausible mechanism for an emerging QSO to directly affect 
the star formation. Although selection effects may play a role, this lends weight to the idea that FeLoBALs and ULIRGs are 
linked in some way. 

It is plausible therefore that the Fe absorption seen in FeLoBALs arises from iron injected into the ISM by an ongoing 
or recent starburst; marking the FeLoBAL phenomenon as a transition stage in a ULIRG when the starburst is at 
or near its end, and the central QSO is starting to throw off its dust cocoon. In this letter we examine the validity of 
this speculation, by observing a sample of nine FeLoBAL QSOs using the Multiband Imaging Photometer for Spitzer 
\citep{rie04} onboard the {\it Spitzer} space telescope \citep{wer04}. We assume a spatially flat cosmology, with 
$H_{0}=70$ km s$^{-1}$ Mpc$^{-1}$, $\Omega=1$, and $\Omega_{\Lambda}=0.7$. Unless otherwise stated, the term 
`IR luminosity' refers to the total luminosity integrated over 1-1000$\mu$m in the rest-frame. 

\section{Observations}
Due to their rarity, assembling a homogenous sample of FeLoBAL QSOs is not trivial. At the time of writing this paper 
a few hundred FeLoBALs are known \citep{tru06}, but when the proposal was being written only 40 or so FeLoBAL QSOs were known 
over the whole sky, most of which lay at $z>1$. We therefore aimed to select a sample that spanned a relatively narrow redshift 
range, and gave a reasonable sampling of the properties of FeLoBAL QSOs, at the expense of homogeneity, with the broader goal 
of obtaining a qualitative idea of the range of their infrared properties. We imposed an upper redshift cut of $z=1.8$ and a 
lower redshift cut of z=1.0. We then randomly selected nine objects. Six of these objects were found via spectroscopic 
observations of the Sloan Digital Sky Survey \citep{hall02}, one (ISO J0056-2738) was discovered serendipitously from followup 
of distant clusters that had been surveyed with ISO \citep{duc02}, one (FBQS J1427+2709) was discovered during spectroscopic 
followup of radio loud quasar candidates from the FIRST survey \citep{bec97}, and one is the `archetype' FeLoBAL QSO LBQS 0059-2735 
\citep{haz87}. The full sample is listed in Table \ref{sample}.

Seven of the sample were observed in July 2006, one (SDSS J0338+0056) was observed in February 2007, and one (LBQS 0059-2735) was observed 
in March 2004, using MIPS at 24$\mu$m, 70$\mu$m and 160$\mu$m (PIDs 30299 \& 82). Following observation, the raw data were 
processed automatically with the MIPS data reduction pipeline at the Spitzer Science Center, which performs standard tasks such 
as image coaddition, sky and dark subtraction, and bias removal. We inspected the output frames from this pipeline, and 
determined that they were of sufficient quality for reasonably accurate ($\sim10\%$) photometry of our targets, hence no further 
reduction steps were performed. Photometry was carried out using the `digiphot' package within the Image Reduction Analysis 
Facility (IRAF) software. We used apertures of 2.45, 2.0 and 2.5 pixels at 24$\mu$m, 70$\mu$m and 160$\mu$m to measure the 
fluxes of our objects, resulting in aperture corrections of 1.698, 3.900 and 3.215 respectively.

\section{Results}
The MIPS fluxes are presented in Table \ref{sample}, along with any available archival IR photometry. Four objects are detected 
in all three MIPS bands, though two of these four are detected only weakly at 160$\mu$m. Three objects are detected at 24$\mu$m 
and $70\mu$m. The remaining two objects are detected only at 24$\mu$m. 

In the absence of IR spectroscopy, detailed measurements of the properties of IR-luminous AGN or starbursts in our sample are 
not possible. We can however constrain both the total IR luminosities, and the contribution from star formation and/or an AGN, 
by fitting the IR photometry simultaneously with the library of model spectral energy distributions (SEDs) for the emission from 
a starburst \citep{efs00} and an AGN \citep{rr95}, following the methods described in \citet{far03}. These model libraries span 
a large number of free parameters (e.g. torus opening angle and line of sight for the AGN, burst lifetime and UV opacity for the 
starbursts) but we here use the complete model libraries solely to estimate the likely range in total, starburst and AGN 
luminosities that are consistent with the available data. The results are presented in Table \ref{sedfits}.

\begin{figure*}
\begin{minipage}{180mm}
\includegraphics[angle=90,width=81mm]{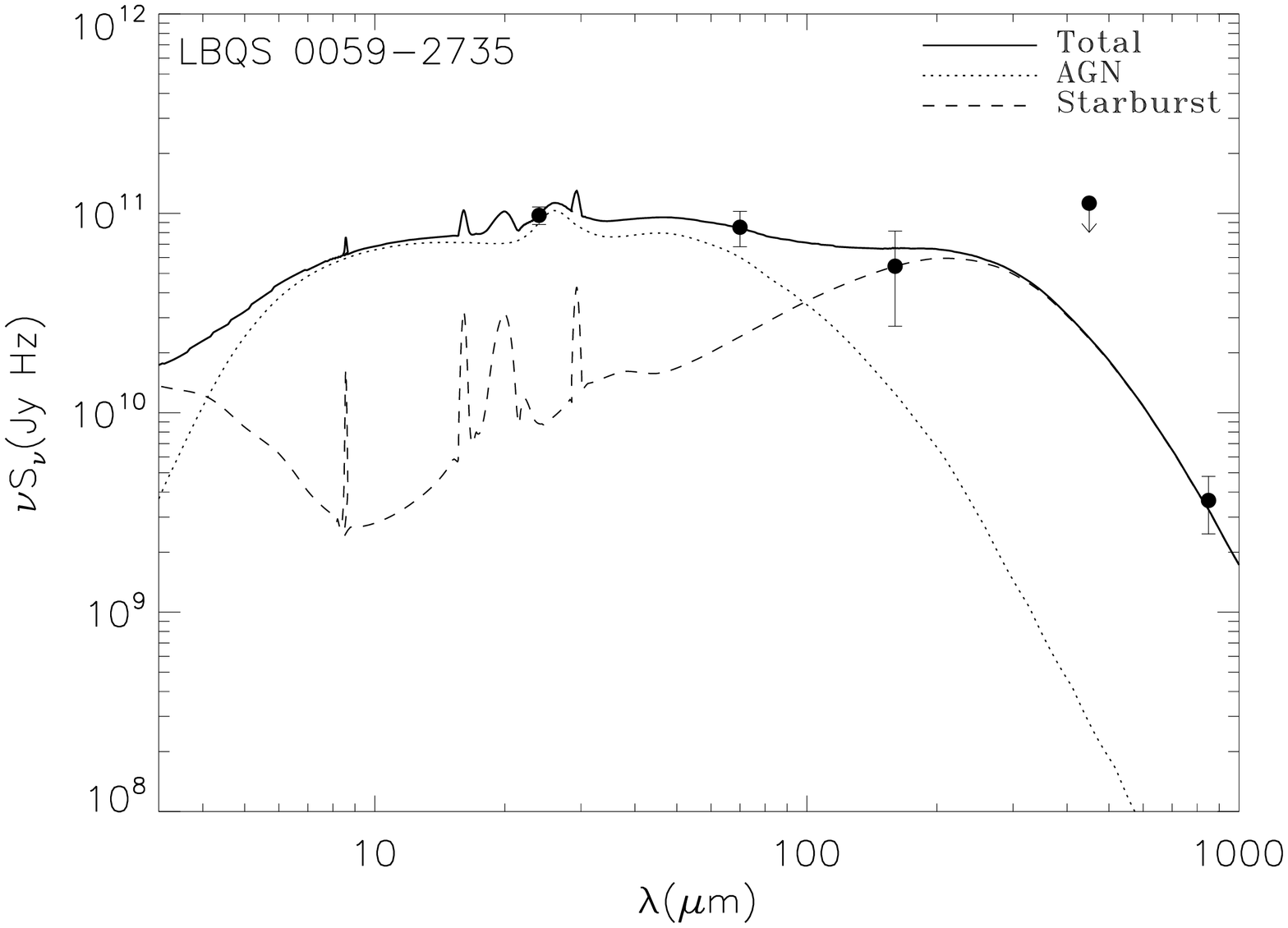}
\includegraphics[angle=90,width=81mm]{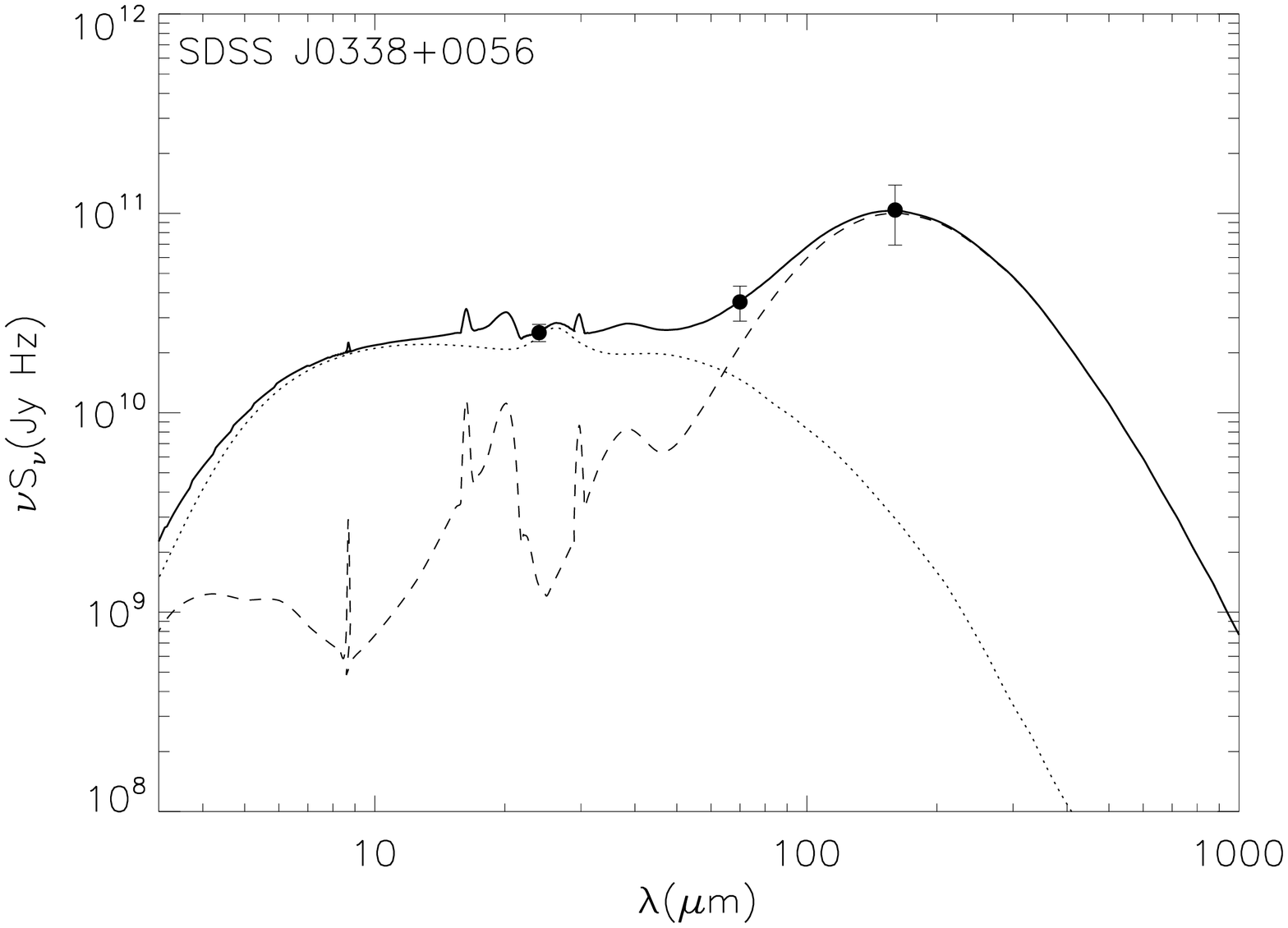}
\end{minipage}
\begin{minipage}{180mm}
\includegraphics[angle=90,width=81mm]{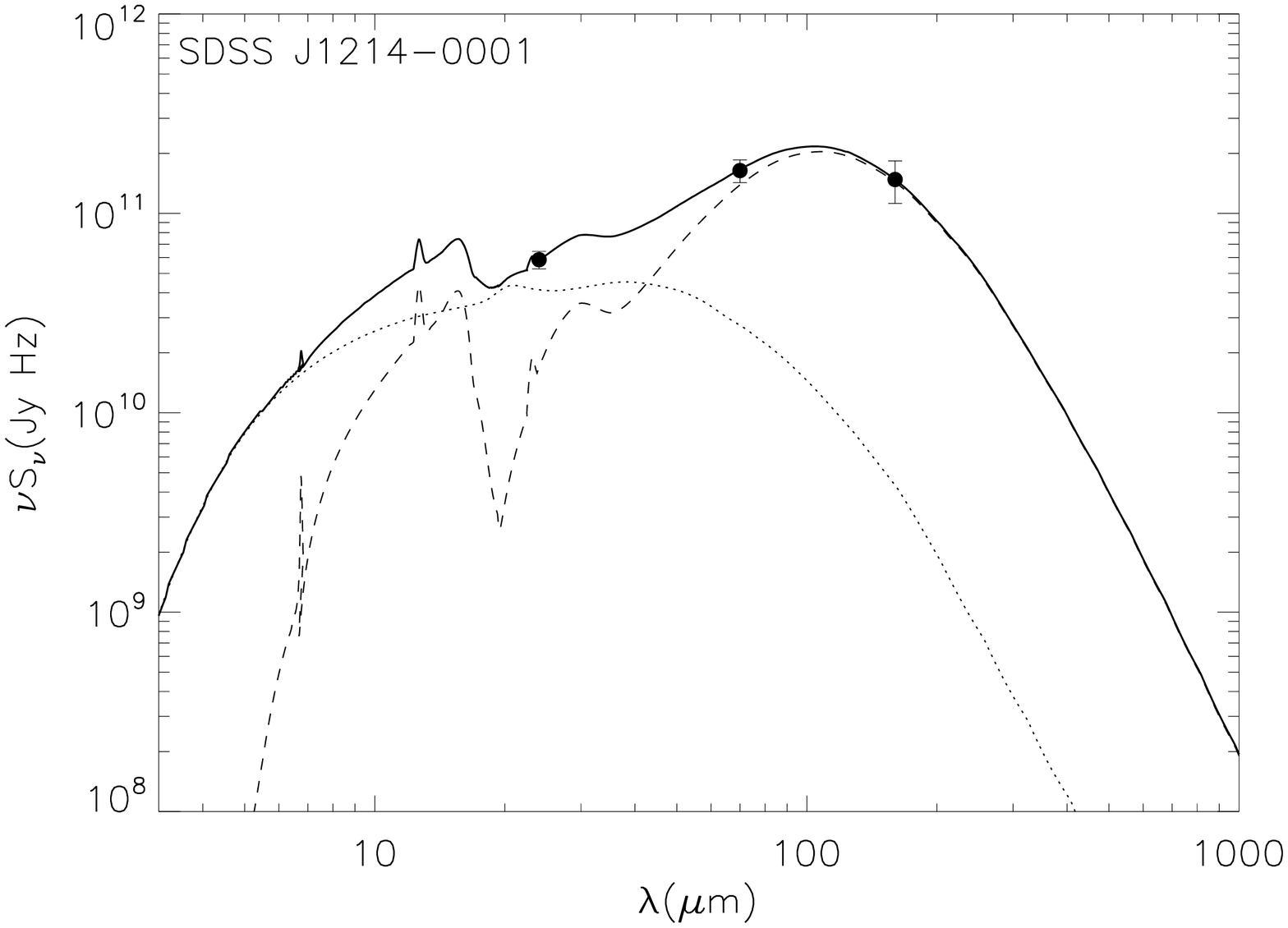}
\includegraphics[angle=90,width=81mm]{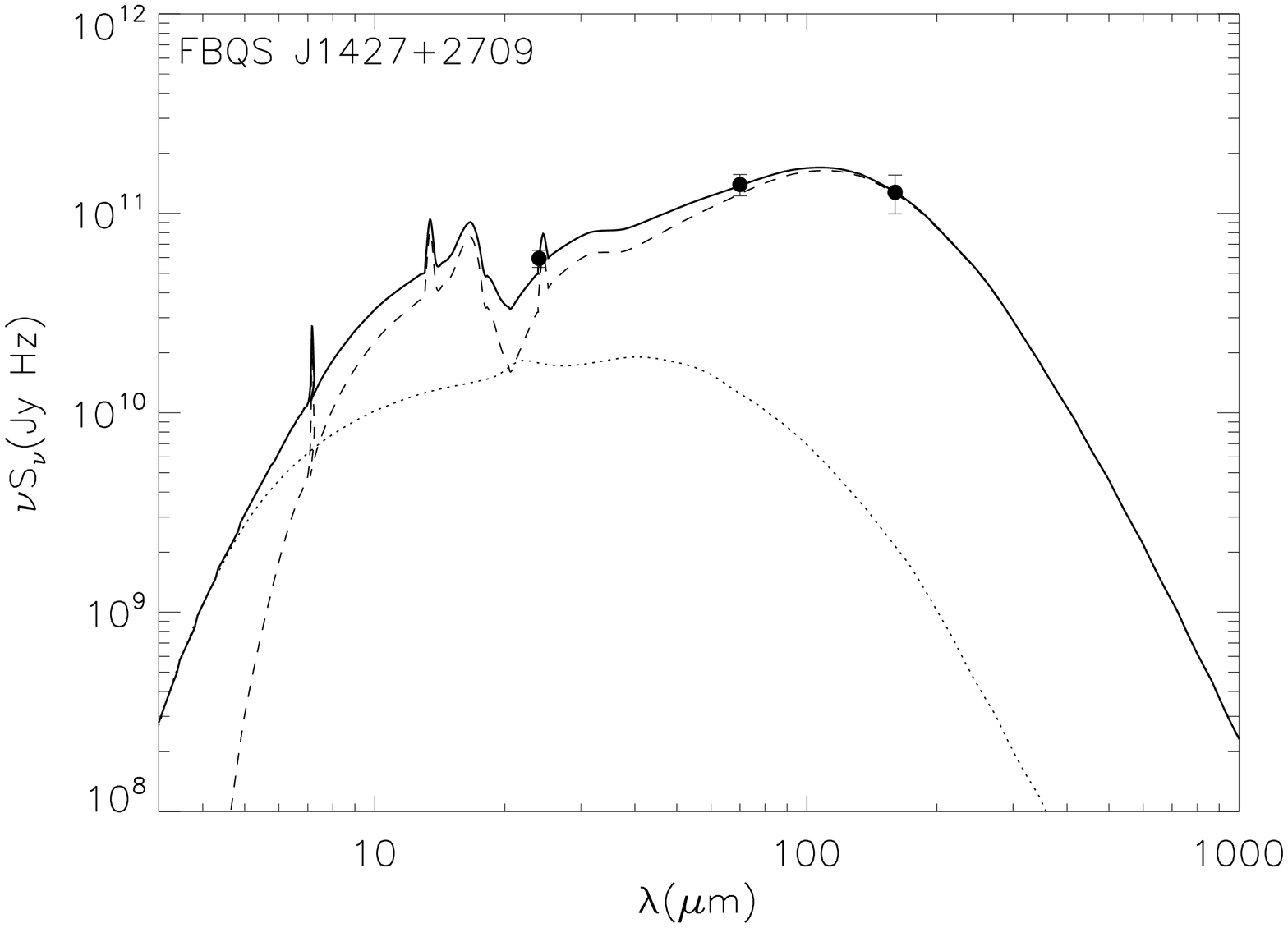}
\end{minipage}
\caption{Best-fit observed-frame SEDs for the four objects with detections in all three MIPS bands. 
The best fits have a starburst and AGN component in all four cases, though 
an AGN contribution to the IR emission is required only in LBQS 0059-2735. The fits do not 
provide any constraints on the contribution from a starburst or an AGN in the near/mid-IR, or on 
spectral features (e.g. PAH luminosities). The fits do however illustrate that a starburst is required 
to explain the emission at longer wavelengths, while remaining consistent with the 24$\mu$m and 
70$\mu$m fluxes. Error bars are 1$\sigma$ while upper limits are 3$\sigma$.  \label{bestseds}}
\end{figure*}

\begin{figure*}
\begin{minipage}{180mm}
\includegraphics[angle=90,width=130mm]{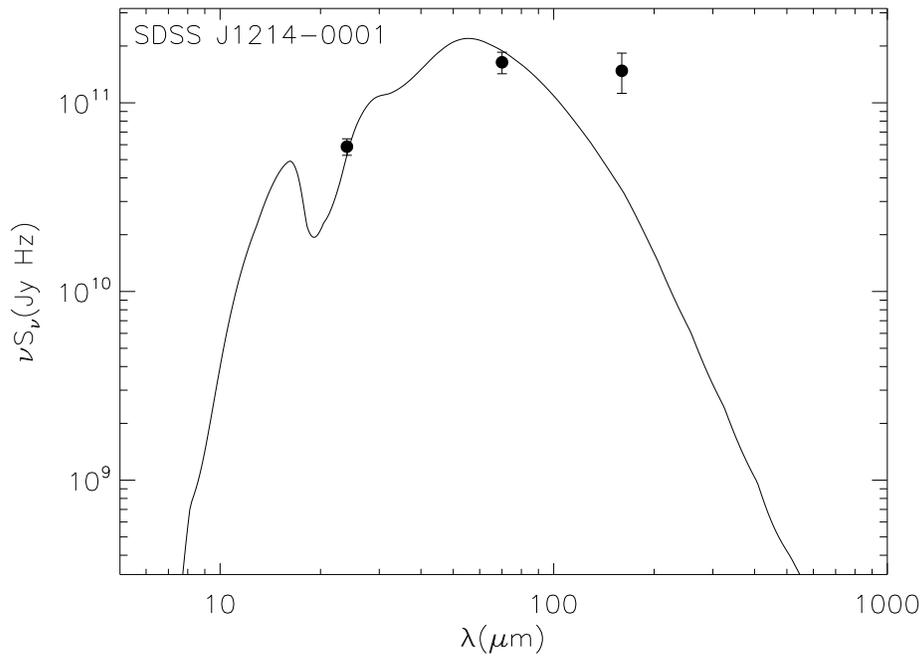}
\end{minipage}
\caption{An example of an observed-frame pure AGN fit to one of the four objects in our sample with detections at 
longer wavelengths, in this case the best possible pure AGN fit to SDSS J1214-0001. Here the AGN model differs 
from that in the lower left panel of Figure \ref{bestseds}; the torus is viewed nearly edge-on, leading to a very high inferred obscuration 
and a strong absorption feature at rest-frame 9.7$\mu$m. Even with such a model however, the fit misses the 
160$\mu$m point by a wide margin.  \label{pureagn}}
\end{figure*}

The four objects with detections in all three MIPS bands are all extremely luminous, with IR luminosities exceeding 
10$^{12.7}$L$_{\odot}$. In all four objects a starburst is required to explain the 160$\mu$m emission while remaining consistent 
with the 24$\mu$m and 70$\mu$m points (for LBQS 0059-2735 an AGN model is consistent with the MIPS data, but the detection at 850$\mu$m 
requires a starburst component). The predicted starburst luminosities exceed 10$^{12.4}$L$_{\odot}$ in all cases, with inferred 
star formation rates of several hundred Solar masses per year. The best fit SEDs for these four objects are given in Figure 
\ref{bestseds}. An example of a pure AGN fit to one of these four objects is shown in Figure \ref{pureagn}. 

The remaining five objects are not detected at 160$\mu$m, hence the constraints on their luminosities and power sources are cruder. 
For these five objects we do not present best-fit SEDs, but merely summarize the results. Three of these five objects are detected 
at 24$\mu$m and 70$\mu$m, and all have IR luminosities exceeding 10$^{12.9}$L$_{\odot}$. The non-detection at 160$\mu$m combined 
with the detection at 70$\mu$m sets an upper limit on any starburst contribution such that an AGN must supply at least some of the 
24$\mu$m and 70$\mu$m emission. Indeed, it is possible to explain the 24$\mu$m and 70$\mu$m emission in all three objects purely 
with an AGN model, though a significant starburst contribution is not ruled out. The final two objects are only detected at 
24$\mu$m. For one of these (ISO J0056-2738) we also have a 15$\mu$m flux from ISO \citep{duc02}, allowing us to conclude that the bulk 
of the IR emission likely arises from an AGN. For SDSS J2336-0107 however, we have only a 24$\mu$m flux, and hence 
cannot set any meaningful constraints on its IR emission.

\section{Discussion}
Starting with the results in this paper, we examine the conjecture that FeLoBAL QSOs represent a specific point in an evolutionary 
sequence between a ULIRG and a `classical' QSO. We initially only consider the total IR luminosities, listed in Table \ref{sedfits}. 
All of our sample are extremely IR-luminous, with luminosities comparable to those of both local ULIRGs \citep{far03,lon06} and to high 
redshift sub-mm bright sources (SMGs, e.g. \citealt{bla02,bor04}). In most cases the upper limits on the IR luminosities exceed 
$10^{13}$L$_{\odot}$, making them potentially comparable in luminosity to high redshift Hyperluminous Infrared Galaxies 
\citep{rr00,far02}. At time of writing there exists no `classical' QSO sample for which direct comparisons to our sample are 
valid (i.e. one at $1.0<z<1.8$, matched in optical magnitude and observed with MIPS), but it is notable that the IR luminosities 
of our sample substantially exceed those of nearly all the Palomar-Green QSOs presented by \citet{haa03}. 

We next consider the power source behind the IR emission. In 5/9 objects, the SED fits predict that the IR emission arises at least 
in part, and possibly entirely, from an AGN, perhaps accompanied by a starburst. In 4/9 objects however, those with detections at 
longer wavelengths, the SED fits demand that a substantial fraction of the total IR emission arises from star formation, with implied 
star formation rates of order a few hundred solar masses per year. This implies that FeLoBAL QSOs are preferentially associated with 
obscured, luminous starbursts that are comparable to the starbursts that are thought to power the majority of the IR emission from 
both local ULIRGs and distant SMGs. There is the caveat that our sample is heterogenous, but if we restrict attention to only the six 
QSOs selected from the SDSS, then the fraction of sources with starbursts, at 2/6, remains equivalent within the errors, though the 
small sample sizes render these conclusions tentative at best. 

A potentially more serious caveat however is that the results in Table \ref{sedfits} are dependent on the assumption that the SED 
libraries used in the fitting span the range in SED shapes of observed IR-luminous starbursts and AGN. This assumption is a 
reasonable one, given that these same SED libraries give good fits to the SEDs of local ULIRGs, and predict starburst and AGN 
luminosities that are consistent with results from other wavelengths \citep{far03}. There is, however, one scenario we cannot 
test for. Although we find that a luminous starburst must be present in the four objects detected at 160$\mu$m, there do exist 
models for the dust distributions around AGN where the dust is so extended (several tens of parsecs) that its outer 
regions are cold enough to emit significantly at wavelengths longward of 100$\mu$m. Such an extended dust distribution could 
account for some or all of the 160$\mu$m emission in these four objects. We cannot formally exclude this scenario, but note 
that interferometric observations of local AGN have found no evidence supporting such extended dust distributions 
(e.g. \citet{jaf04}). We therefore do not consider this possibility to be likely. 

Finally, we consider our results together with those from previous work. Prior evidence that suggests links between FeLoBAL 
QSOs and ULIRGs includes; (1) the only two systems at low redshifts known to contain FeLoBALs are both ULIRGs (e.g. \citealt{far05}), 
(2) FeLoBAL QSOs at high redshifts may be involved in interactions \citep{gre02}, and (3) large-scale winds in FeLoBAL QSOs 
may provide a mechanism for the emerging AGN to affect star formation in the host galaxy \citep{dek02}. To these 
we add (4) FeLoBAL QSOs are, as a class, extremely IR-luminous, with IR luminosities comparable to those of ULIRGs at low and 
high redshifts, and (5) star formation powers a significant fraction of the IR emission in many, and possibly all, FeLoBAL QSOs, 
with implied star formation rates comparable to those inferred for local and high redshift ULIRGs. Overall therefore, our 
results combine with those from previous work to construe strong, albeit indirect evidence for evolutionary links between 
FeLoBAL QSOs and ULIRGs. We therefore propose that FeLoBALs mark galaxies and QSOs in which an extremely luminous starburst 
is approaching its end, and in which a rapidly accreting supermassive black hole is in the last stages of casting off its dust 
cocoon. FeLoBAL signatures in high redshift QSOs and galaxies may thus be an efficient way of selecting sources at a critical 
point in their evolution. 

\acknowledgments
We thank the referee for a very helpful report. This work is based on observations made with the Spitzer Space Telescope, which is operated by the Jet Propulsion
Laboratory, California Institute of Technology under a contract with NASA. Support for this work was provided by NASA.
This research has made extensive use of the NASA/IPAC Extragalactic Database (NED) which is operated by the Jet Propulsion
Laboratory, California Institute of Technology, under contract with NASA. JA acknowledges support from the 
Science and Technology Foundation (Portugal) through the research grant POCI/CTE-AST/58027/2004. RSP thanks the University of Hertfordshire for
support.

\begin{deluxetable}{lcccccccccc}
\tabletypesize{\scriptsize}
\tablecolumns{11}
\tablewidth{0pc}
\tablecaption{FeLoBAL QSO Sample, and infrared photometry \label{sample}}
\tablehead{
\colhead{Galaxy}&\colhead{RA (J2000)}&\colhead{Dec}&\colhead{z}&\colhead{$m_{i}$}&\colhead{$f_{15}$}&\colhead{$f_{24}$}&\colhead{$f_{70}$}&\colhead{$f_{160}$}&\colhead{$f_{450}$\tablenotemark{a}}&\colhead{$f_{850}$\tablenotemark{a}}
}
\startdata
ISO  J005645.1-273816     & 00 56 45.15 & -27 38 15.6 & 1.78 & 20.95\tablenotemark{f}  & 1.33\tablenotemark{b} & 1.56  & $<$7.0 & $<$50                    & --       & --  \\
LBQS 0059-2735            & 01 02 17.02 & -27 19 48.8 & 1.59 & 17.39\tablenotemark{f}  & --                    & 7.83  &   19.9 &    29.0\tablenotemark{d} & $<$168.9 & 10.3 \\
SDSS J033810.85+005617.6  & 03 38 10.85 & +00 56 17.6 & 1.63 & 18.33                   & --                    & 2.02  &    8.4 &    55.5\tablenotemark{e} & --       & -- \\
SDSS J115436.60+030006.3  & 11 54 36.60 & +03 00 06.4 & 1.46 & 17.74                   & --                    & 7.59  &   17.9 & $<$50                    & --       & -- \\
SDSS J121441.42-000137.8  & 12 14 41.43 & -00 01 37.9 & 1.05 & 18.77                   & --                    & 4.69  &   38.3 &    78.9                  & --       &  --\\
FBQS J142703.6+270940     & 14 27 03.64 & +27 09 40.3 & 1.17 & 18.11                   & --                    & 4.76  &   32.6 &    68.1                  & --       & -- \\
SDSS J155633.78+351757.3  & 15 56 33.80 & +35 17 58.0 & 1.50 & 18.01                   & 8.30\tablenotemark{c} & 13.93 &   24.7 & $<$50                    & $<$86.1  & $<$3.6 \\
SDSS J221511.93-004549.9  & 22 15 11.94 & -00 45 49.9 & 1.48 & 16.49                   & --                    & 10.4  &   27.2 & $<$50                    & --       & -- \\
SDSS J233646.20-010732.6B & 23 36 45.10 & -01 07 32.4 & 1.29 & 21.73                   & --                    & 0.48  & $<$7.0 & $<$50                    & --       & -- \\
\enddata
\tablecomments{All fluxes are quoted in mJy. The `small' field size was used for all three channels, using the default pixel scale at 70$\mu$m. Exposure times per cycle and number of cycles for all sources except LBQS 0059-2735 were 30s, 10s, 10s and 5, 7 and 4 respectively, giving on-source exposure times of 2260s, 755s and 85s. For LBQS 0059-2735 the exposure times were shorter; 3s, 10s and 10s with 1, 1, and 3 cycles respectively. Errors are typically 10\% on the 24$\mu$m fluxes, 20\% on the 70$\mu$m fluxes, and 25\% on the 160$\mu$m fluxes. Limits are 3$\sigma$.}
\tablenotetext{a}{\citet{lew03}}\tablenotetext{b}{\citet{duc02}}\tablenotetext{c}{\citet{cla98}}\tablenotetext{d}{2$\sigma$ detection}
\tablenotetext{e}{2.7$\sigma$ detection}\tablenotetext{f}{R band magnitude}

\end{deluxetable}

\begin{deluxetable}{lcccc}
\tabletypesize{\scriptsize}
\tablecolumns{8}
\tablewidth{0pc}
\tablecaption{Infrared Luminosities \label{sedfits}}
\tablehead{
\colhead{Galaxy}&\colhead{L$_{tot}$}&\colhead{L$_{AGN}$}&\colhead{L$_{SB}$}
}
\startdata
ISO J0056-2738  & 12.48-13.00 &    12.40-13.00  & $<$12.90                  \\
LBQS 0059-2735  & 12.95-13.20 &    12.80-13.00  &    12.44-12.82            \\
SDSS J0338+0056 & 12.90-13.40 & $<$12.40        &    12.60-13.40            \\
SDSS J1154+0300 & 12.89-13.10 &    12.70-12.95  & $<$12.98                  \\
SDSS J1214-0001 & 12.70-12.88 & $<$12.44        &    12.68-12.84            \\
FBQS J1427+2709 & 12.81-13.01 & $<$12.70        &    12.70-13.00            \\
SDSS J1556+3517 & 13.10-13.30 &    12.90-13.23  & $<$12.50                  \\
SDSS J2215-0045 & 13.00-13.40 &    12.50-13.40  & $<$12.80                  \\
SDSS J2336-0107 & 11.70-12.90 &     --          &  --                       \\
\enddata
\tablecomments{Quoted luminosities are the logarithm of the rest-frame 1-1000$\mu$m luminosity, in units of solar luminosities 
(3.826$\times10^{26}$ Watts), derived from the SED fits. Limits and ranges are 3$\sigma$.}
\end{deluxetable}

\clearpage

\begin{thebibliography}{}

\bibitem[Arav et al.(2001)]{ara01} 
Arav, N., et al.\ 2001, \apj, 561, 118 

\bibitem[Becker et al.(1997)]{bec97} 
Becker, R.~H., Gregg, M.~D., Hook, I.~M., McMahon, R.~G., White, R.~L., \& Helfand, D.~J.\ 1997, 
\apjl, 479, L93 

\bibitem[Blain et al.(2002)]{bla02} 
Blain, A.~W., Smail, I., Ivison, R.~J., Kneib, J.-P., \& Frayer, D.~T.\ 2002, \physrep, 369, 111 

\bibitem[Borys et al.(2004)]{bor04} 
Borys, C., Scott, D., Chapman, S., Halpern, M., Nandra, K., \& Pope, A.\ 2004, \mnras, 355, 485 

\bibitem[Branch et al.(2002)]{bra02} 
Branch, D., Leighly, K.~M., Thomas, R.~C., \& Baron, E.\ 2002, \apjl, 578, L37

\bibitem[Canalizo \& Stockton(2001)]{can01} 
Canalizo, G., \& Stockton, A.\ 2001, \apj, 555, 719 

\bibitem[Clavel(1998)]{cla98} 
Clavel, J.\ 1998, \aap, 331, 853 

\bibitem[Duc et al.(2002)]{duc02} 
Duc, P.-A., et al.\ 2002, \aap, 389, L47 

\bibitem[Efstathiou et al.(2000)]{efs00} 
Efstathiou, A., Rowan-Robinson, M., \& Siebenmorgen, R.\ 2000, \mnras, 313, 734 

\bibitem[Elvis(2000)]{elv00} 
Elvis, M.\ 2000, \apj, 545, 63

\bibitem[Farrah et al.(2002)]{far02} 
Farrah, D., Serjeant, S., Efstathiou, A., Rowan-Robinson, M., \& Verma, A.\ 2002, \mnras, 335, 
1163 

\bibitem[Farrah et al.(2003)]{far03} 
Farrah, D., Afonso, J., Efstathiou, A., Rowan-Robinson, M., Fox, M., \& Clements, D.\ 2003, \mnras, 
343, 585 

\bibitem[Farrah et al.(2005)]{far05} 
Farrah, D., Surace, J.~A., Veilleux, S., Sanders, D.~B., \& Vacca, W.~D.\ 2005, \apj, 626, 70 

\bibitem[Gregg et al.(2002)]{gre02} 
Gregg, M.~D., Becker, R.~H., White, R.~L., Richards, G.~T., Chaffee, F.~H., \& Fan, X.\ 2002, 
\apjl, 573, L85 

\bibitem[Haas et al.(2003)]{haa03} 
Haas, M., et al.\ 2003, \aap, 402, 87 

\bibitem[Hall et al.(2002)]{hall02} 
Hall, P.~B., et al.\ 2002, \apjs, 141, 267 

\bibitem[Hazard et al.(1987)]{haz87} 
Hazard, C., McMahon, R.~G., Webb, J.~K., \& Morton, D.~C.\ 1987, \apj, 323, 263 

\bibitem[Hutsem{\'e}kers \& Lamy(2000)]{hun00} 
Hutsem{\'e}kers, D., \& Lamy, H.\ 2000, \aap, 358, 835 

\bibitem[Jaffe et al.(2004)]{jaf04} 
Jaffe, W., et al.\ 2004, \nat, 429, 47

\bibitem[Kawakatu et al.(2006)]{kaw06} 
Kawakatu, N., Anabuki, N., Nagao, T., Umemura, M., \& Nakagawa, T.\ 2006, \apj, 637, 104 

\bibitem[de Kool et al.(2002)]{dek02} 
de Kool, M., Becker, R.~H., Arav, N., Gregg, M.~D., \& White, R.~L.\ 2002, \apj, 570, 514 

\bibitem[Lacy et al.(2002)]{lac02} 
Lacy, M., Gregg, M., Becker, R.~H., White, R.~L., Glikman, E., Helfand, D., \& Winn, J.~N.\ 
2002, \aj, 123, 2925 

\bibitem[Lewis et al.(2003)]{lew03} 
Lewis, G.~F., Chapman, S.~C., \& Kuncic, Z.\ 2003, \apjl, 596, L35 

\bibitem[Lipari et al.(1994)]{lip94} 
Lipari, S., Colina, L., \& Macchetto, F.\ 1994, \apj, 427, 174 

\bibitem[Lonsdale et al.(2006)]{lon06} 
Lonsdale, C.~J., Farrah, D., \& Smith, H.~E.\ 2006, Astrophysics Update 2, 285, astroph 0603031 

\bibitem[Lynds(1967)]{lyn67} 
Lynds, C.~R.\ 1967, \apj, 147, 837 

\bibitem[Murray \& Chiang(1998)]{mc98} 
Murray, N., \& Chiang, J.\ 1998, \apj, 494, 125 

\bibitem[Ogle et al.(1999)]{ogl99} 
Ogle, P.~M., Cohen, M.~H., Miller, J.~S., Tran, H.~D., Goodrich, R.~W., \& Martel, A.~R.\ 1999, \apjs, 
125, 1 

\bibitem[Priddey et al.(2007)]{pri07} 
Priddey, R.~S., Gallagher, S.~C., Isaak, K.~G., Sharp, R.~G., McMahon, R.~G., \& Butner, 
H.~M.\ 2007, \mnras, 374, 867 

\bibitem[Proga et al.(2000)]{pro00} 
Proga, D., Stone, J.~M., \& Kallman, T.~R.\ 2000, \apj, 543, 686 

\bibitem[Reichard et al.(2003)]{rei03} 
Reichard, T.~A., et al.\ 2003, \aj, 126, 2594 

\bibitem[Rieke et al.(2004)]{rie04} 
Rieke, G.~H., et al.\ 2004, \apjs, 154, 25 

\bibitem[Rowan-Robinson(1995)]{rr95} 
Rowan-Robinson, M.\ 1995, \mnras, 272, 737 

\bibitem[Rowan-Robinson(2000)]{rr00} 
Rowan-Robinson, M.\ 2000, \mnras, 316, 885 

\bibitem[Sanders et al.(1988)]{san88} 
Sanders, D.~B., Soifer, B.~T., Elias, J.~H., Madore, B.~F., Matthews, K., Neugebauer, G., \& 
Scoville, N.~Z.\ 1988, \apj, 325, 74 

\bibitem[Schmidt \& Hines(1999)]{sch99} 
Schmidt, G.~D., \& Hines, D.~C.\ 1999, \apj, 512, 125 

\bibitem[Sprayberry \& Foltz(1992)]{spr92} 
Sprayberry, D., \& Foltz, C.~B.\ 1992, \apj, 390, 39

\bibitem[Trump et al.(2006)]{tru06} 
Trump, J.~R., et al.\ 2006, \apjs, 165, 1 

\bibitem[Voit et al.(1993)]{voi93} 
Voit, G.~M., Weymann, R.~J., \& Korista, K.~T.\ 1993, \apj, 413, 95 

\bibitem[Werner et al.(2004)]{wer04} 
Werner, M.~W., et al.\ 2004, \apjs, 154, 1 

\bibitem[Weymann et al.(1991)]{wey91} 
Weymann, R.~J., Morris, S.~L., Foltz, C.~B., \& Hewett, P.~C.\ 1991, \apj, 373, 23 

\bibitem[Williams et al.(1999)]{wil99} 
Williams, R.~J.~R., Baker, A.~C., \& Perry, J.~J.\ 1999, \mnras, 310, 913 

\bibitem[Willott et al.(2003)]{will03} 
Willott, C.~J., Rawlings, S., \& Grimes, J.~A.\ 2003, \apj, 598, 909 

\end{thebibliography}
\end{document}